\PassOptionsToPackage{protrusion, expansion}{microtype}

\documentclass[conference, nofonttune]{IEEEtran}

\usepackage{pgfplotstable}
\usepackage{pgfplots}

\usepackage[maxbibnames=99]{biblatex}

\usepackage{comment}
\usepackage{amsmath, amsthm, amssymb, mathtools, bm} 
    
\usepackage{graphicx, xcolor}               
    \definecolor{B}    {HTML}{2b66d3}
    \definecolor{B2}   {HTML}{003399}
    \definecolor{Bv}   {HTML}{0000EB}   
    \definecolor{R}    {HTML}{c9171e}
    \definecolor{R2}   {HTML}{d7003a}
    \definecolor{INK}  {HTML}{595857}
    \definecolor{Y}    {HTML}{f1c40f}
    \definecolor{G}    {HTML}{009a00}
    \definecolor{GRAY} {HTML}{808080}
    \definecolor{MAUVE}{HTML}{9400D1}
\usepackage{url, hyperref}                  
	\hypersetup{
	   pdfborder={0 0 0}}
\usepackage{adjustbox}                      
\usepackage{booktabs, multirow}             
\usepackage{caption, subcaption}            
\usepackage{enumitem}                       
\usepackage{lipsum}                         
\usepackage{setspace}                       
\usepackage[boxed]{algorithm}       		
    \floatstyle{plaintop}
\usepackage{algpseudocode}                  
    \algrenewcommand{\alglinenumber}[1]{{\scriptsize\bfseries\ttfamily\color{R}#1}}
    \algrenewcommand\algorithmiccomment[1]{\hfill{\color{gray}\fontsize{6.5}{7}\selectfont\sffamily$\triangleright$\ #1}}
    
\makeatletter
\newcommand\fs@ruled@nomiddle{\def\@fs@cfont{\bfseries}\let\@fs@capt\floatc@ruled
    \def\@fs@pre{}%
    \def\@fs@mid{\vskip 1pt\hrule height.8pt depth0pt \kern2pt}%
    \def\@fs@post{\vskip 5pt\hrule height.8pt depth0pt \relax}%
  \let\@fs@iftopcapt\iftrue}
\renewcommand\fst@algorithm{\fs@ruled@nomiddle}
\makeatother

\usepackage{xpatch}
\makeatletter
    \xpatchcmd{\algorithmic}{\ALG@tlm\z@}{\ALG@tlm\z@\leftmargin 10pt}{}{}
\makeatother

\usepackage{footnote}                       
\usepackage{listings}						
\usepackage{colortbl}
\usepackage{tikz}
\usetikzlibrary{positioning}

\usepackage{array}
\usepackage{rotating}

\newcommand{\BOLD}{\fontfamily{ugq}\selectfont}

\newcommand{\TABLETITLE}{\BOLD}
\newcommand{\TABLECAPTION}{\color{gray}\scriptsize}

\newcommand{\tableSETUP}{\centering\footnotesize\fontfamily{cmr}\selectfont}

\newcommand{\SPEEDUP}{\color{gray}\bfseries}

\newcommand{\func}[1]{{\ttfamily#1}}

\hypersetup{pdfauthor={C. Rivera}, pdftitle={Optimizing Huffman Decoding}}

\newcommand{\SEC}{{\S}}

\newcommand{\cody}[1]{{\color{black}#1}}

\usepackage[scale=.97]{tgheros} 
\usepackage[scale=.95]{newtxtt}

\usepackage[normalem]{ulem}

\IEEEoverridecommandlockouts

\begin{document}

\title{Optimizing Huffman Decoding for Error-Bounded Lossy Compression on GPUs}

\newcommand{\BetterMark}{$^\star$}

\author{
Cody Rivera\BetterMark,
Sheng Di\IEEEauthorrefmark{3},
Jiannan Tian\IEEEauthorrefmark{2},
Xiaodong Yu\IEEEauthorrefmark{3},
Dingwen Tao\IEEEauthorrefmark{2}\BetterMark\thanks{Corresponding author: Dingwen Tao (\url{dingwen.tao@wsu.edu}), School of EECS, Washington State University, Pullman, WA 99164, USA.},
Franck Cappello\IEEEauthorrefmark{3}\\
\BetterMark%
Department of Computer Science, University of Alabama, Tuscaloosa, AL, USA\\
\IEEEauthorrefmark{2}%
School of Electrical Engineering and Computer Science, Washington State University, Pullman, WA, USA\\
\IEEEauthorrefmark{3}%
Mathematics and Computer Science Division, Argonne National Laboratory, Lemont, IL, USA
}

\maketitle

\thispagestyle{plain}
\pagestyle{plain}

\begin{abstract}
    More and more HPC applications require fast and effective compression techniques to handle  large volumes of data in storage and transmission. Not only do these applications need to compress the data effectively during simulation, but they also need to perform decompression efficiently for post hoc analysis. SZ is an error-bounded lossy compressor for scientific data, and cuSZ is a version of SZ designed to take advantage of the GPU's power. At present, cuSZ's compression performance has been optimized significantly while its decompression still suffers considerably lower performance because of its sophisticated lossless compression step---a customized Huffman decoding. In this work, we aim to significantly improve the Huffman decoding performance for cuSZ, thus improving the overall decompression performance in turn. To this end, we first investigate two state-of-the-art GPU Huffman decoders in depth.
    Then, we propose a deep architectural optimization for both algorithms. Specifically, we take full advantage of CUDA GPU architectures by using shared memory on decoding/writing phases, online tuning the amount of shared memory to use, improving memory access patterns, and reducing warp divergence. Finally, we evaluate our optimized decoders on an Nvidia V100 GPU using \cody{eight} representative scientific datasets. Our new decoding solution obtains an average speedup of \cody{3.64}$\times$ over cuSZ's Huffman decoder and improves its overall decompression performance by \cody{2.43}$\times$ on average.
\end{abstract}



\section{Introduction}
\label{sec:intro}

High-performance computing (HPC) applications are generating increasingly large amounts of data. For example, Hardware/Hybrid Accelerated Cosmology Code (HACC) is a cosmological simulation package designed for HPC environments. For HACC simulations consisting of trillions of particles, one snapshot of the simulation takes up 220 TB of data, while an entire simulation run can take up 22 PB of data~\cite{hacc}. Another such application is the pruning and compression of deep neural networks, which are becoming deeper and wider, and therefore larger, as computing capacity is increasing and deep neural networks are becoming more widely used~\cite{deepsz19}. However, leading supercomputers such as Summit~\cite{summit} have limited storage capacities of approximately 50$\sim$200 PB to share between hundreds of users. Thus, these applications require data reduction techniques that attain both high performance and high compression ratios. SZ, for instance, is a lossy data compressor that aims to achieve these goals. It offers over a 2x increase in compression ratio over state-of-the-art compressors. Furthermore, it allows users to specify how much error they wish to tolerate in their data and allows them to make a tradeoff between data distortion and compression ratios~\cite{sz17}.

As a result of the evolution of supercomputer architecture (e.g., more powerful GPUs on a single node), many HPC applications are being implemented on graphics processing units (GPUs) due to their high performance and parallelism. For example, a recent study of cosmological simulations on the Summit supercomputer~\cite{hacc-summit} shows a significant performance improvement compared to prior work without using GPUs~\cite{habib2016hacc}.
However, even ignoring the time to transfer the uncompressed data from the GPU to the CPU, CPU-based lossy compressors would still cause more than 10\% overhead of the overall performance, which would limit the I/O performance gain by lossy compression, according to prior studies~\cite{sz17, jin2020understanding}.
Thus, several lossy compressor development teams have recently released their GPU versions to reduce the compression overhead. These GPU versions can both accelerate the compression computation and reduce the time needed to transfer the data between the GPU and CPU 
after the compression.
For example, both SZ and MGARD~\cite{ainsworth2017mgard} have a GPU adaptation (known as cuSZ~\cite{cusz2020} and cuMGARD~\cite{chen2021accelerating}), which have been implemented for GPU hardware, and more specifically Nvidia’s CUDA platform.

\cody{Furthermore, an important use case of lossy compression is to reduce the memory footprint by storing the data in the compressed format and calculating from the lossy data in memory \cite{use-case-Franck}.
Compared to the amount of data handled in extreme-scale applications, memory is scarce on HPC systems. Lossy compression can be introduced to ease this pressure in applications that can tolerate some loss of fidelity in their working data. An example of in-memory compression is Wu et al.'s work on quantum circuit simulation \cite{wu2019quantum}, where in-memory compression reduced the total memory usage on 4,096 nodes from 32 exabytes to 768 terabyte. This compression allowed for 2$\sim$16 more qubits to be simulated than if no compression was used.
Another example is Jin et al.'s work on reverse time migration (RTM) simulation \cite{jin2021improving}, where in-memory compression reduced the memory usage by about 10$\times$ on average.
Moreover, for some applications, in-memory compression can decrease repetitive computations and accelerate execution \cite{use-case-Franck}. For example, Gok et al.'s work on quantum chemistry simulation---GAMESS \cite{pastri}---proposes to calculate, compress, and write each unique data block of two-electron integrals into the memory once; and whenever a block is needed again in simulations, it is read from the memory and decompressed. Compared with the original GAMESS, where all blocks are generated and consumed by the simulation on the fly and are then deleted
from the memory, Gok et al.'s approach achieves a reduction in the block re-computation costs. Note that all these computations should be done at runtime because integral blocks are generated and consumed repeatedly during a simulation. Consequently, in-memory (de)compression throughputs are critical to the overall performance. 
To this end, in this work, we focus on improving decompression throughputs on GPUs without considering GPU-to-CPU data-movement overheads, as decompressed data do not need to be transferred to CPUs or disks.
}

An important component of both cuSZ and cuMGARD is Huffman coding, a classic lossless compression technique initially developed by David Huffman in 1952~\cite{Huffman-original}. Tian~\textit{et al.}'s work proposes an optimized Huffman encoder for GPUs\cite{tian2021revisiting}; their work has been applied to improve cuSZ's compression throughput.
Although efficient compression is important to speedup the overall data movement, efficient decompression is also important to enable fast and effective post-analysis based on compressed data. However, Huffman decoders used by error-bounded lossy compressors currently employ only a limited degree of parallelism and do not fully exploit the GPU's power. Two state-of-the-art works propose improved Huffman decoding on the GPU: one of these works, Weißenberger and Schmidt's~\cite{weissenberger2018massively}, uses the self-synchronization property of Huffman codes to extract greater parallelism, while the other work, Yamamoto~\textit{et al.}'s~\cite{yamamoto2020gaparray}, proposes a new data structure called a gap array to extract greater parallelism. But both works suffer from two main issues: (1) they do not fully take advantage of the GPU architecture for performance optimization, and (2) they must be adapted for use in error-bounded lossy compression. 

To facilitate their efficient use in scientific data compression, we explore 
both Huffman decoding algorithms in depth. Furthermore, we identify opportunities for deep optimization of both algorithms based on GPU architecture considerations. Finally, we adapt both algorithms for use in error-bounded lossy compression
such as cuSZ, and then evaluate them on \cody{eight} representative scientific datasets. The contributions of our work are summarized as follows:
\begin{itemize}[noitemsep, topsep=2pt, leftmargin=1.3em]
    \item We analyze Weißenberger and Schmidt's and Yamamoto~\textit{et al.}'s algorithms in depth by evaluating their performance on scientific datasets and understanding their tradeoffs.
    \item  We perform a deep architectural optimization for both Huffman decoding algorithms by using shared memory in the decoding and writing phase, improving memory access patterns, and reducing warp divergence.
    \item We propose an efficient approach to online tune the amount of shared memory used to decode different parts of the data based on the data characteristics.
    \item We adapt our optimized decoders to multi-byte data for cuSZ and evaluate
          them on \cody{eight} scientific datasets.
          Experiments show our solution can improve decoding throughput
          by \cody{3.64}$\times$, compared with cuSZ's na\"ive decoder, and can improve cuSZ's overall performance by \cody{2.43}$\times$, on average.
\end{itemize}

In \SEC\ref{sec:bg}, we present background information about scientific data compression, Huffman coding, and GPU-based lossy compression.
In \SEC\ref{sec:pa}, we discuss both Weißenberger and Schmidt's and Yamamoto~\textit{et al.}'s Huffman decoding algorithms in detail, comparing them and discussing their limitations.
In \SEC\ref{sec:design}, we describe our architectural optimizations for efficient Huffman decoding, as well as our adaptations to enable decoding of scientific datasets.
In \SEC\ref{sec:eval}, we show the experimental evaluation results on scientific datasets.
In \SEC\ref{sec:related} and \SEC\ref{sec:conclusion}, we discuss the related work and conclude our work.

\section{Background}
\label{sec:bg}

\subsection{Scientific Data Compression}\label{sub:back-comp}
Scientific data compression has been studied for decades and categorized into two types of compression: lossless compression and lossy compression.
Lossless compressors such as FPZIP~\cite{lindstrom2006fast} and FPC~\cite{fpc} keep the data intact but can only provide a compression ratio of about 2$\times$ on scientific data~\cite{son2014data}.
Lossy compression, on the other hand, can compress data beyond lossless compression (typically one or more orders of magnitude) but introduces information loss in the reconstructed data.
In recent years, a new generation of high accuracy lossy compressors for scientific data have been proposed and developed for scientific floating-point data, such as SZ~\cite{sz16, sz17, sz18}, ZFP~\cite{zfp}, and MGARD~\cite{ainsworth2017mgard}. These lossy compressors provide parameters that allow users to finely control the information loss introduced by lossy compression. 
Unlike traditional lossy compressors such as JPEG~\cite{jpeg} for images (in integers), SZ, ZFP, and MGARD are designed to compress floating-point data and can provide a strict error-controlling scheme based on the user's requirements.
Generally, lossy compressors provide multiple compression modes, such as error-bounding mode and fixed-rate mode.
Error-bounding mode requires users to set an error type, such as the point-wise absolute error bound and point-wise relative error bound, and an error bound level (e.g., $10^{-3}$). The compressor ensures that the differences between the original data and the reconstructed data do not exceed the user-set error bound level.
Fixed-rate mode means that users can set a target bit-rate (the number of bits to represent each data point), and the compressor guarantees the actual bit-rate of the compressed data to be lower than the user-set value.

\subsection{Huffman Coding}\label{sub:back-huff}
Huffman Coding is a classic technique developed by David Huffman in 1952 for performing lossless compression~\cite{Huffman-original}. It encodes a fixed-length value as a variable-length code. We call the fixed-length input value an \textit{input symbol}, and we call the variable-length output value a \textit{codeword}. In Huffman coding, space savings result from the fact that more frequently occurring symbols are represented by codewords with fewer bits, and vice versa for less frequently occurring symbols. Huffman codewords are also prefix-free; no one codeword can be a prefix of any other codeword.

\subsection{CUDA Architecture}

\textit{Thread}: The thread is the basic programmable unit that allows programmers to use  massive numbers of CUDA cores. CUDA threads are grouped at different levels including warp, block, and grid.

\textit{Warp}: The warp is a basic-level scheduling unit in CUDA associated with SIMD (single-instruction multiple-data). Specifically, the threads in a warp achieve convergence when executing exactly the same instruction; otherwise, warp divergence happens. In the current
CUDA architecture, the number of threads in a warp is 32, hence, it works as 32-way SIMT when converging. However, when diverging happens,
diverged threads add extra overhead to the execution \cite{xiang2014warp}. 

\textit{Block}: Unlike the warp, the thread block (or simply block) is a less hardware-coupled description of thread organization, as it is explicitly seen in the kernel configuration when launching one. Threads in the same block can access the shared memory, a small pool of fast programmable cache. On one hand, shared memory is bound to active threads, which are completely scheduled by the GPU hardware; however, on the other hand, a grid of threads may exceed the hardware-supported number of active threads at a time. 
As a result, the data stored in the shared memory used by the previous batch of active threads may be invalid when the current or following batch of active threads are executing. 
Thus, we must carefully tune the shared memory size for different workloads toward high performance.

\textit{Grid}: A grid encompasses the entire set of blocks that are launched as part of a CUDA kernel. 
Usually, the grid of threads describes either the entire problem or a working set of the problem at hand.
Moreover, all the blocks within a grid share a common configuration; each block within a grid contains the same amount of shared memory and the same number of threads per block.

\subsection{Error-bounded Lossy Compression on GPU}\label{sub:back-gpu}

SZ, ZFP, and MGARD were first developed for CPU architectures, and all started rolling out their GPU-based lossy compression recently.
The SZ team, the ZFP team, and the MGARD team released their CUDA versions, called cuSZ~\cite{cusz2020}, cuZFP~\cite{cuZFP}, and cuMGARD~\cite{chen2021accelerating}, respectively.
All the versions provide much higher throughputs for compression and decompression compared with their CPU versions~\cite{cusz2020, jin2020understanding, tian2021optimizing}.
Compared with cuSZ and cuMGARD, cuZFP provides slightly higher compression throughput, but it only supports fixed-rate mode~\cite{jin2020understanding}, limiting its adoption in practice.
Both cuSZ and cuMGARD use Huffman encoding to achieve high compression ratios and their decompression throughput is greatly limited by slow Huffman decoding on GPUs, but cuSZ has a much higher throughput than cuMGARD~\cite{tian2021optimizing, chen2021accelerating}.
Thus, in this work, we focus on optimizing Huffman decoding for cuSZ.

\section{Analysis of Existing Huffman Decoders for Error-Bounded Lossy Compression}\label{sec:pa}

\subsection{Coarse-Grained Versus Fine-Grained Parallelism}\label{sub:grain-par}

The current implementation of parallel Huffman decoding in cuSZ requires a number of fixed-size chunks containing thousands of codewords to be decoded sequentially by many threads~\cite{cusz2020}. Such a solution is called a coarse-grained solution, as there are fewer threads performing a relatively large amount of work. Although such a solution may provide good performance on a multi-core CPU, as multi-core CPUs tend to have either tens or hundreds of powerful,  \cody{independent} cores, GPUs \cody{have thousands of interdependent cores that work best when running together in lock-step. Since GPU cores are interdependent, and parallel Huffman decoding is not particularly amenable to lock-step parallelism, the apparent performance of a single thread is relatively weak compared to a CPU thread. Nevertheless, since GPUs have so many cores, we can improve performance by proposing a fine-grained solution---a solution that launches many threads that operate on fewer data elements.} 

It is possible to extract greater parallelism from cuSZ's existing coarse-grained Huffman decoder by decreasing the size of each chunk; however, since Huffman codes are variable length, very small chunks may not be able to be filled, leaving empty space in chunks that degrade the compression ratio. One avenue for increasing the parallelism in Huffman decoding is to determine a starting point in the bitstream for each thread; two connected strategies for doing this are described in the following subsections.  

\subsection{Self-Synchronization Based Huffman Decoding}\label{sub:ss-dec}
Weißenberger and Schmidt proposed a parallel algorithm for Huffman decoding on the GPU using a property of Huffman codes called \textsc{self-synchronization}~\cite{weissenberger2018massively}. Their technique is in turn based on an earlier CPU-based parallel Huffman decoder by Klein and Wiseman~\cite{klein2003parhuff}. This algorithm is designed to work on pure Huffman codes; no modifications to the Huffman encoding step need to be done. It uses the self-synchronization property of Huffman codes to determine where in the bitstream each thread starts decoding, allowing for finer-grained parallelism than a chunk-based approach.

\subsubsection{Self-Synchronization}\label{subsub:self-sync}


The self-synchronization property of Huffman codes is the tendency for a Huffman decoder to correct itself even if a few bits of the input were skipped in error~\cite{ferguson1984sync}. An example of self-synchronization is as follows: consider the bit pattern \texttt{``111000010111000''} with the Huffman codebook from~\cite{ferguson1984sync}, shown in Listing~\ref{tab:self-sync-ex}. A correct decoding of this pattern is \texttt{``(11)(10)(00)(010)(11)(10)(00)''}, or \texttt{``CBADCBA''}. However, if one bit is skipped in the input, then the pattern is decoded as \texttt{``(11)(00)(00)(10)(11)(10)(00)''}, or \texttt{``\underline{\textbf{CAAB}}CBA''}. The first four characters' outputs are incorrect, but after decoding four erroneous characters (8 bits), the decoder starts decoding the correct characters, i.e., it self-synchronizes. There is a possibility that for a given Huffman codebook, there are codestreams that never self-synchronize, but Klein and Wiseman \cite{klein2003parhuff} demonstrated that for practical datasets, self-synchronization was achieved in less than 72 bits on average. 

\begin{minipage}{\linewidth}
\begin{lstlisting}[numbers=none, escapeinside={(*}{*)}, caption={Example self-synchronizing codebook from~\cite{ferguson1984sync}.},captionpos=b, label={tab:self-sync-ex}, breaklines=true]
symbol (*$\to$*) codeword = {
    'A': '00', 'B': '10', 
    'C': '11', 'D': '010', 'E': '011' }
\end{lstlisting}
\end{minipage}

\subsubsection{Fast Decoding via Self-Synchronization}\label{subsub:ss-fast-dec}

\begin{figure}
    \centering\fontfamily{lmr}\selectfont
    \resizebox{\linewidth}{!}{\begin{tikzpicture}[y=-1cm]

\begin{scope}[yshift=2cm]

\node[font=\fontfamily{lmr}\selectfont\footnotesize] at (1-.25, 0){
\begin{tabular}{@{} c c @{}}
\bfseries \begin{tabular}{@{}c@{}}source\\[-.6ex]letter\end{tabular} & codeword \\[1ex]
A & 00  \\
B & 10  \\
C & 11  \\
D & 010 \\
E & 011 \\
\end{tabular}
};

\node[] at (6,0) {
\footnotesize
\begin{tabular}{@{} rl @{}}
\raisebox{-1ex}{
\begin{tikzpicture}\draw[latex-latex, thin] (0,-.2) -- (0,.2);\end{tikzpicture}
}
  & synchronization point \\
\fontfamily{ugq}\selectfont\tiny OK & verified sync. point \\
& \bfseries\color{red} Thread0 \color{blue} Thread1 \\
& \bfseries\color{gray} Thread2 \color{black} Thread3
\end{tabular}
};

\end{scope}

\begin{scope}
\node[font=\scriptsize\bfseries, text width=3em, align=right] at (-1.25,0) {subseq. 0,1};
\foreach \i/\sym in { 0/1, 1/0, 2/0, 3/0, 4/1, 5/1, 6/0, 7/0, 8/1, 9/1, 10/1, 11/1, 12/1, 13/0, 14/0, 15/1 } {
  \node[font=\footnotesize\fontfamily{lmr}\selectfont] at (\i*0.5, 0) {\sym}; }
\node [draw, minimum width=8cm, minimum height=.4cm] at (4 -.25,0){};

\draw(4-.25, -.2) -- (4-.25, .2);

\begin{scope}[]
\draw[blue, latex-latex, ultra thick] (4-.25, -.4) -- (4-.25, .4);
\node[font=\tiny\fontfamily{ugq}\selectfont] at (4-.25, .55) {OK};
\end{scope}

\begin{scope}[xshift=-4cm]
\draw[red, latex-latex, ultra thick] (4-.25, -.4) -- (4-.25, .4);
\node[font=\tiny\fontfamily{ugq}\selectfont] at (4-.25, .55) {OK};
\end{scope}

\begin{scope}[xshift=3cm]
\draw[gray, latex-latex, ultra thick] (4-.25, -.4) -- (4-.25, .4);
\node[font=\tiny\fontfamily{ugq}\selectfont] at (4-.25, .55) {OK};
\end{scope}

\end{scope}

\begin{scope}[yshift=-1.25cm]
\node[font=\scriptsize\bfseries, text width=3em, align=right] at (-1.25,0) {subseq. 2,3};
\foreach \i/\sym in { 0/0, 1/1, 2/0, 3/0, 4/0, 5/0, 6/0, 7/0, 8/1, 9/1, 10/1, 11/0, 12/1, 13/0, 14/0, 15/0 } {
  \node[font=\footnotesize\fontfamily{lmr}\selectfont] at (\i*0.5 , 0) {\sym}; }
\node [draw, minimum width=8cm, minimum height=.4cm] at (4 -.25,0){};

\draw(4-.25, -.2) -- (4-.25, .2);

\begin{scope}[xshift=-.5cm]
\draw[latex-latex, ultra thick] (4-.25, -.4) -- (4-.25, .4);
\node[font=\tiny\fontfamily{ugq}\selectfont] at (4-.25, .55) {OK};
\end{scope}

\end{scope}
    
\end{tikzpicture}}
    \caption{Illustrating the final step of the self-synchronization phase of Weißenberger and Schmidt's decoder}
    \label{fig:selfsync-illustrate}
\end{figure}
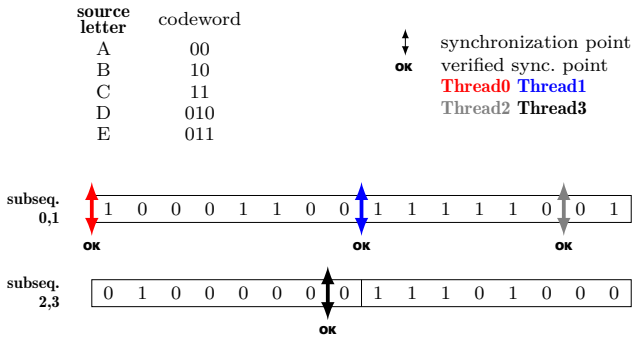

Although self-synchronization is most often examined in the context of error correction, it can be used to determine starting locations for many-threaded Huffman decoders. This is the basic technique used in Weißenberger and Schmidt's decoder~\cite{weissenberger2018massively}. To determine where each thread starts, first, each thread is placed at evenly-spaced intervals throughout the Huffman bitstream. Then, each thread decodes but does not write, a certain number of bits in the Huffman bitstream, and stores the location where it stopped decoding, called a synchronization point. This decoding gives each thread an opportunity to self-synchronize; if this is the case, the stored location points to a valid codeword. Each thread's synchronization point is validated by the previous thread. If the previous thread `meets' up with the current thread's synchronization point, then the synchronization point refers to a valid codeword; otherwise, the threads continue decoding until a valid synchronization point is found. For example, Figure \ref{fig:selfsync-illustrate} illustrates the conclusion of the self-synchronization phase of the algorithm, when each thread has had its synchronization point verified; if a thread starts decoding at one of these points, it will decode valid characters in the encoded string \texttt{``BACACCBDBAAEBBA''}.

An overview of Weißenberger and Schmidt's algorithm is described as follows:
\begin{itemize}[noitemsep, topsep=1pt, leftmargin=1.3em]
    \item Use self-synchronization to determine the synchronization points within each sequence;
    \item Use self-synchronization to adjust the synchronization points between sequences;
    \item Use a prefix sum to determine where each thread writes to in the output array;
    \item Have each thread decode data, starting at a synchronization point, and write it to the output array.
\end{itemize}
To understand the steps of the above algorithm, we define a \textsc{sequence} to be a chunk of the input data that a single CUDA thread block operates on. A subsequence is a subdivision of a sequence that a single CUDA thread works on. A subsequence in turn is divided into \textsc{units}: unsigned 32-bit numbers that contain the individual codewords. The number of synchronization points within a given dataset is equal to the number of subsequences in the input data. Steps 1 and 2 ensure that all the synchronization points reference valid codewords. During this process, the number of valid codewords in each subsequence is recorded. These numbers are then prefix-summed in step 3 to determine the first output index for each subsequence. Finally, in step 4, each thread starts decoding at its synchronization point and outputs data starting at the index generated by the prefix sum. We refer readers to~\cite{weissenberger2018massively} for more details.

\textbf{Limitation:}
Although the self-synchronization based Huffman decoding
allows for finer-grained Huffman decoding, 
it contains some major performance bottlenecks. One major bottleneck is determining synchronization points; to do this, the algorithm needs to decode the input data multiple times depending on the particular dataset to be decoded.

\subsection{Gap Arrays}\label{sub:gap-array}
To address the self-synchronization's performance issue, Yamamoto~\textit{et al.} work proposed a new data structure called a gap array to eliminate this decoding bottleneck, in exchange for some extra encoding overhead~\cite{yamamoto2020gaparray}.
Note that although the work also proposes an optimized encoding scheme, our main focus is its decoding scheme. 
Similar to Weißenberger and Schmidt's algorithm, Yamamoto~\textit{et al.}'s decoder divides its input data into sequences, subsequences, and units as defined above. However, instead of finding out where each thread starts decoding within a subsequence by determining synchronization points, this information is stored alongside the compressed data in a gap array. A \textsc{gap array} is a byte array, with one byte per subsequence, that indicates to each thread how many bytes it must skip before accurate data can be decoded. 
For example, a gap array for the codewords in Figure \ref{fig:selfsync-illustrate}
would be $[0, 0, -2, -1]$, as these are the offsets from the subsequence boundaries that each thread would have to keep track of in order to decode correctly.
The gap array is used in combination with a technique called Single Kernel Soft Synchronization (SKSS) to determine output indices, decode the codewords, and write output symbols to memory. We refer readers to~\cite{yamamoto2020gaparray} for more details.

\textbf{Limitation:} Although the gap array removes the necessity of performing the self-synchronization phase and speeds up the decoding, gap arrays introduce other overheads. These overheads include the extra space required to store the gap array as well as extra work for the encoder. Nevertheless, the extra space required to store a gap array is minimal, as Yamamoto~\textit{et al.} has shown that the size of the gap array is less than 3\% of the size of the data on their tested datasets of varying compression ratios~\cite{yamamoto2020gaparray}. However, the extra work the encoder must perform in generating the gap array means that the encoder and the decoder must be coupled. This reduces the flexibility of gap-array-based Huffman decoding, as it will not be able to decode Huffman codes generated by encoders not designed to create gap arrays. Nevertheless, since there are compelling reasons to use both self-synchronization and gap arrays in practice (will be discussed in detail in \SEC \ref{sub:use-case}), we consider both solutions in our optimization work.

\subsection{Challenges of Using Existing Huffman Decoders}\label{sub:perf-issues}

\begin{figure}[]
     \centering\sffamily\footnotesize
     \includegraphics[trim={.1in 0 0 0}, width=\linewidth]{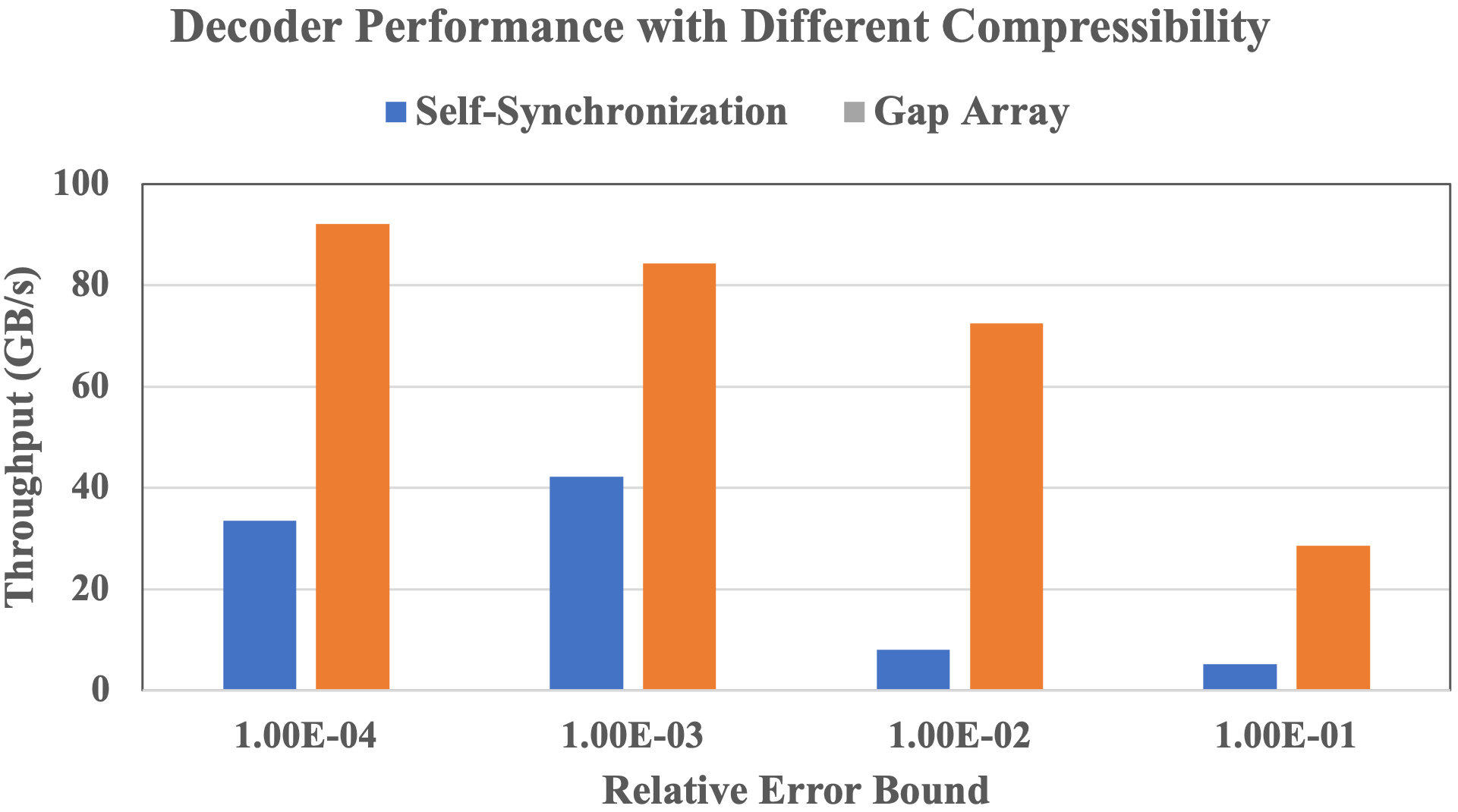}
    \caption{Decoding performance versus error bounds on HACC dataset. Note that the larger the error bound, the larger the compression ratio.}
     \label{fig:high-cr-motiv}
\end{figure}

Both Weißenberger and Schmidt's and Yamamoto~\textit{et al.}'s works have been evaluated across a wide range of general-purpose datasets~\cite{weissenberger2018massively,yamamoto2020gaparray}. However, in error-bounded lossy compression such as cuSZ, since input data are quantization codes, the resulting quantization codes are often highly compressible, especially in a well-predicted dataset.
\cody{Our experiments show that} 
both their works underperform on high-compressible datasets, as can be seen in Figure \ref{fig:high-cr-motiv}. Note that in general, as error bound increases, the resulting quantization codes become easier to compress. The figure illustrates a drop in the throughput of both decoders as data becomes more easily compressible. Thus, not only do we optimize both their solutions in general, but we also specifically focus on techniques to optimize high compression-ratio cases commonly seen in scientific data reduction.

\section{Design Methodology}
\label{sec:design}

To allow efficient Huffman decoding of multi-byte input such as quantization codes in cuSZ, 
we perform a series of architectural optimizations on both Weißenberger and Schmidt's and Yamamoto~\textit{et al.}'s solutions. We start with Weißenberger and Schmidt's implementation as a baseline and adapt it to multi-byte input. 
We continue by examining the bottlenecks in the current algorithm: (1) the intra-sequence synchronization phase (i.e., Step 1), for self-synchronization based Huffman decoding; and (2) the decoding and writing phase (i.e., Step 4), for both self-synchronization and gap-array-based Huffman decoding. \cody{Details of the decoders can be found both in this subsection as well as in our online repository\footnote{\url{https://github.com/codyjrivera/ipdps22-opthuffdec}.}.}

\subsection{Optimized Self-Synchronization}\label{sub:ss-optim}

We note that although the average behavior of self-synchronization is well-predictable, the amount of data each thread needs to decode to achieve self-synchronization can vary, as aforementioned. 
More severely, although each thread needs to decode only two subsequences on average to find and validate a synchronization point\footnote{Note that one subsequence containing four 32-bit units results in 128 bits decoded per subsequence.}, up to 5\% of threads decode greater than two subsequences, and individual threads can decode up to 125 subsequences on the test datasets in order to find and validate synchronization points. 
As a result, the local unpredictability of self-synchronization hinders GPU implementation because if one thread decodes more subsequences than other threads in the same CUDA warp, the other threads are held up until the longest-running thread finishes its job. This inefficiency is exacerbated by the fact that, in the self-synchronization phase, a CUDA block-wide thread barrier is required for correctness. Thus, the longest-running thread determines the running time of the entire block; and other threads within a block remain idle.

A potential solution to this issue is to perform load balancing to ensure that threads in a block are not idle; however, due to the overhead of load balancing and the relatively low occurrence of exceptionally long-running threads, we do not pursue the load balancing approach. 
Instead, we do optimize the intra-sequence self-synchronization kernel to minimize the impact of long-running threads and conform more closely to the CUDA architecture. 
Specifically, in the original code for self-synchronization based Huffman decoding, after each thread discovers and validates a synchronization point, it busy-waits not only until the longest-running thread in the thread block finishes but also until the maximum possible number of subsequences that a thread may decode until self-synchronization is reached (e.g., 128 subsequences in this case). To allow thread blocks to exit as early as possible, we record each thread's ``finished'' status in a Boolean variable. By using the CUDA warp primitive \texttt{\_\_all\_sync}, we can determine
whether all the threads have finished finding their synchronization points; and if so, we will terminate the kernel immediately,
freeing up warps within a CUDA Streaming Multiprocessor to execute other blocks in the kernel. 
This optimized intra-sequence self-synchronization runs, on average, \cody{11\%} faster than the baseline code, and these benefits are concentrated in lower compression ratio datasets, in which this phase is a more significant bottleneck than in high compression-ratio datasets.

\subsection{Optimized Decoding and Writing of Codewords}\label{sub:decode-write}

In both presented decoders, threads decode and write codewords directly to the GPU's global memory. There is a stride between different threads' output indices; this stride reflects the number of codewords that can be found between the two threads' input locations. This counters one of the characteristics of CUDA's memory architecture: coalesced memory loads and stores. Specifically, a coalesced memory access is when sequential global memory transactions are combined with each other to reduce the number of memory transactions actually performed. For example, in CUDA, a 32-thread warp writing 32-bit values to sequential locations in memory 
have its write requests processed as a single 128-byte transaction. Note that inefficient memory access patterns result in many more transactions being made, which decreases the throughput of the global memory. 

For high compression-ratio datasets, this memory inefficiency is even worse, because not only are the gaps between adjacent threads' output indices large, but also the number of values written to global memory, and hence the number of transactions are large. This is a significant factor in the dramatic drop in performance with high compression-ratio datasets shown in Figure \ref{fig:high-cr-motiv}. Although Yamamoto~\textit{et al.}'s work has each thread write multiple symbols at a time to global memory~\cite{yamamoto2020gaparray}, performance still eventually degrades at high ratios, as can be seen in the same figure.  

\begin{algorithm}
\centering
    \newcommand{\varhere}[1]{{\text{\fontfamily{lmr}\selectfont\textit{#1}}}}
\newcommand{\funchere}[1]{\textsc{\bfseries#1}}
\newcommand{\algoCommentFont}{\fontsize{6.5}{7}\selectfont\color{gray}\sffamily}
\newcommand{\IN}{\textcolor{R}{\bfseries in}}

\scriptsize\ttfamily
\begin{algorithmic}[1] 
\vspace{.5ex}
\Statex {\sffamily\color{R}$\bullet$\bfseries\small\ \func{DecodeWrite} --- decode and write using shared memory}
\vspace{.5ex}
\State sharedBuffer[n] \Comment{The shared memory buffer of size $n$}
\State si <- outIndex[blockIdx.x $\cdot$ blockDim.x]
\State ei <- outIndex[(blockIdx.x + 1) $\cdot$ blockDim.x]
\State gid <- threadIdx.x + blockDim.x $\cdot$ threadIdx.x

\State tempEnd <- ei
\While{si < ei}
    \State start <- outIndex[gid] - si, end <- outIndex[gid + 1]
    \If {si $\leq$ start \textbf{\color{R}and} end $\leq$ si + n }
        \State outArray[start $\ldots$ end) <- \funchere{Decode}(inArray, startPoint[gid])
        \Statex \Comment{If symbols can fit into the buffer, decode them}
    \ElsIf {start < si + n \textbf{\color{R}and} end $\geq$ si + n }
        \State tempEnd <- outIndex[gid] 
        \Statex \Comment{Executed by one thread if buffer is not large enough; results in another iteration}
    \EndIf
    \State outArray[si $\ldots$ tempEnd) = sharedBuffer[0 $\ldots$ tempEnd - si)
    \Statex \Comment{This write is performed cooperatively by threads in the block}
    \State si <- tempEnd
\EndWhile
\end{algorithmic}
    \caption{Decoding and writing using a shared memory buffer.}\label{algo:decode-write}
\end{algorithm}

To solve this issue, we propose to first decode the input data into a thread block-local buffer, and then write it out sequentially to global memory to attain coalesced and hence efficient writes. For the thread block-local buffer, we use shared memory. Specifically, first, given the global output index of each thread, the kernel will compute the local index where each thread will put the decoded symbol within the shared memory buffer. Then, the kernel will decode and have each decoder write its data into the shared memory. Finally, all threads in the thread block cooperatively write the data in the shared memory to the global memory output array. Note that the codebook that is used for decoding is kept in global memory; since this codebook is shared across all thread blocks, it is kept in cache, so we do not need to consider keeping a codebook in shared memory and can dedicate the shared memory for the decoding buffer. Note that if the shared memory is not large enough to store all the data that threads inside the block will decode, that the shared memory will be filled up with the initial chunk of decoded data by the initial group of threads, that data will be written, and then the rest of the threads will fill up the shared memory with the rest of the decoded data. More details about this procedure can be found in Algorithm \ref{algo:decode-write}.

\subsection{Shared Memory Tuning for Decoding and Writing}\label{sub:shmem-tuning}
\begin{figure}[]
    \centering\sffamily\footnotesize
    \includegraphics[trim={.1in 0 0 0}, width=\linewidth]{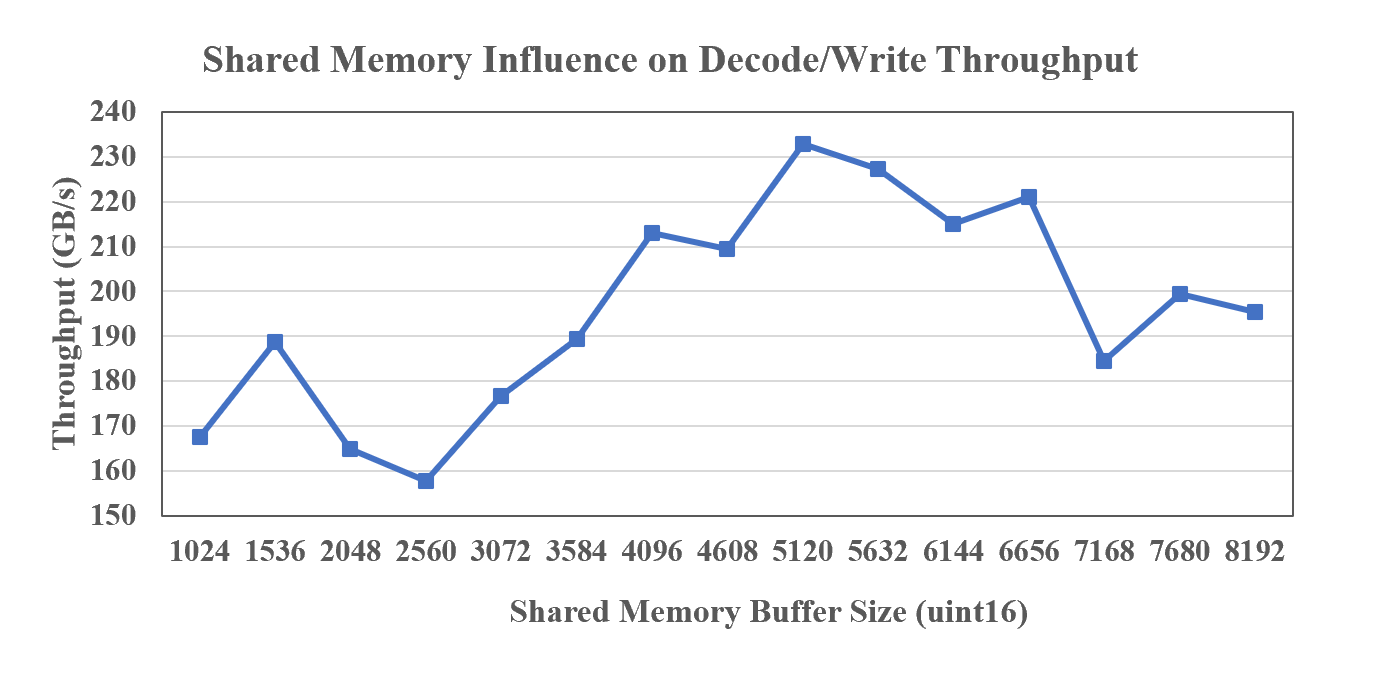}

    \caption{\cody{Throughput of the decoding and writing phase with different shared memory sizes on the quantization codes generated by cuSZ on HACC dataset with relative error bound $10^{-3}$.}}
    \label{fig:tune-motiv}
\end{figure}

The method proposed for decoding the codewords and writing the decoded symbols back to memory requires a certain amount of shared memory to be allocated to the appropriate kernel. Choosing this amount of shared memory can significantly impact the performance of this phase of the decoder, because (1) allocating too little shared memory can reduce parallelism, and (2) allocating too much shared memory can reduce occupancy. 
This can be illustrated in Figure~\ref{fig:tune-motiv}, evaluated on our HACC dataset described in \SEC~\ref{sub:eval-setup}, with an error bound of $10^{-3}$. Note that the difference between the lowest and highest throughput is around \cody{32\% (i.e., 233 GB/s vs. 158 GB/s).}

A first approximation to the amount of shared memory allocated would be to allocate an amount of shared memory proportional to the compression ratio: in this case, the compression ratio is 3.86, and the corresponding buffer size, rounding up 256 spots, would be 4096 16-bit integers. However, the optimal size of the shared memory buffer, in this case, is 5120. Therefore, an improved strategy of allocating shared memory is needed.

To this end, we propose a strategy that more effectively reflects the characteristics of actual data, where some portions of the data are highly-compressible but others are not so highly-compressible. We launch separate kernels for decompressing input sequences with different compression ratios. We determine which sequence is to be decoded by each kernel using an online selection process. That way, each sequence is decoded by a kernel launched with an optimal amount of shared memory.

\begin{algorithm}
\centering
    \newcommand{\varhere}[1]{{\text{\fontfamily{lmr}\selectfont\textit{#1}}}}
\newcommand{\funchere}[1]{\textsc{\bfseries#1}}
\newcommand{\algoCommentFont}{\fontsize{6.5}{7}\selectfont\color{gray}\sffamily}
\newcommand{\IN}{\textcolor{R}{\bfseries in}}

\scriptsize\ttfamily
\begin{algorithmic}[1] 
\vspace{.5ex}
\Statex {\sffamily\color{R}$\bullet$\bfseries\small\ \func{ShmemOptDecodeWrite} --- decode and write with optimal shared memory use}
\vspace{.5ex}
\color{red} 
\State compRatio[n] \Comment{The \cody{precomputed} compression ratios of the $n$ sequences}
\ForAll {i \IN \ [0 \ldots n) concurrently}
    \State compClass[i] <- \funchere{ClassifyCR}(compRatio[i])
\EndFor \Comment{Classifies all the compression ratios into one of the $T_{high} + 1$ groups}
\State compClassFreq <- \funchere{ParHistogram}(compClass)
\Statex \Comment{Finds out how many sequences fall into each group}
\State compIndex <- [0, 1, \ldots, n - 1]
\State \funchere{ParKeyValueSort}(compClass, compIndex)
\Statex \Comment{Allows decoding kernels to access sequences in its compression ratio group}
\color{blue}
\State compClassStart[0] <- 0
\For {i \IN \ [1 \ldots T\textsubscript{high} + 1)}
    \State compClassStart[i] <- compClassStart[i - 1] + compClassFreq[i - 1]
\EndFor \Comment{Determines where in \texttt{compIndex} each compression ratio group starts}
\ForAll {i \IN \ [0 \ldots T\textsubscript{high} + 1) asynchronously}
    \State {\color{red} \funchere{DecodeWrite}(optShmem(i), compClassStart[i], compClassFreq[i])}
\EndFor \Comment{Launches decoding kernels for each of the compression ratio groups. Each kernel gets an amount of shared memory optimized for its compression ratio group.}

\end{algorithmic}
    \caption{Our proposed shared memory optimization that partitions input sequences among kernels launched with different amounts of shared memory. \cody{Lines implemented by host code are in blue, while lines implemented by CUDA kernels are in red.}}\label{algo:shmem-opt}
\end{algorithm}

\cody{We now give more details about this strategy, which are illustrated in Algorithm \ref{algo:shmem-opt}. First, this strategy requires each sequence's compression ratio as input. This is taken from an earlier phase of each algorithm: the self-synchronization phase for Wei{\ss}enberger and Schmidt's algorithm and redundant decoding in Yamamoto et al.'s algorithm required to determine where each thread writes its data. After this is done,} (1) the shared memory optimization starts by classifying each sequence's compression ratio into $T_{high} + 1$ groups, where $T_{high}$ is an architecture-specific threshold\cody{, on lines 2-4}. This classification is then stored inside an \cody{on-device} array. \cody{$T_{high}$} of these groups are the compression ratios in the intervals $(0, 1], (1, 2], ..., (T_{high} - 1, T_{high}]$, and the $T_{high} + 1$-th group encompasses compression ratios in the interval $(T_{high}, 16]$. Thus, at most \cody{$T_{high} + 1$} different kernels with varying amounts of shared memory are launched. (2) The array is then histogrammed \cody{on the GPU}, in order to see how many sequences fall into each compression ratio group. The algorithm used is the same variation of G\'omez-Luna~\textit{et al.}~\cite{gomez2013improving} that is used in cuSZ. (3) Once the classification array is histogrammed\cody{, on line 5}, it is then sorted \cody{on the GPU} as part of a key-value sort, with the classification being used as the key and a sequential list of indices being used as the values. The resulting values will be the primary means by which sequences in a particular compression ratio group are accessed and decoded. The algorithm used is the \texttt{DeviceRadixSort} routine in CUB~\cite{nvidia-cub}. Furthermore, since $T_{high}$ is fairly small, sorting $T_{high} + 1$ groups is fast using CUB (will be proved in Table \ref{tab:eval-default}). (4) \cody{After being transferred back to the CPU, the histogram} is then used to generate a prefix sum that indicates where in the list of indices the sequences belonging to that compression ratio group begin. (5) Finally, up to $T_{high} + 1$ kernels are launched with an amount of shared memory (mostly) proportional to their corresponding compression ratio group's upper bound. For example, sequences in the $(3, 4]$ compression ratio group would be decoded by a kernel with a shared memory buffer of length 4096. \cody{Each kernel is launched on a separate CUDA stream in order to allow the CUDA driver maximum flexibility in scheduling and running the $T_{high} + 1$ kernels.}  These kernels then finish decoding the data and write their data to the output array. 

\begin{table}[ht]
\tableSETUP
    \resizebox{\linewidth}{!}{\begin{tabular}{@{} >{\bfseries\scshape}l rrrrrrrr@{}}
                         & \TABLETITLE HACC & \TABLETITLE EXAALT & \TABLETITLE CESM & \TABLETITLE Nyx  & \TABLETITLE Hurr. & \TABLETITLE \cody{QMC.} & \TABLETITLE \cody{RTM} & \TABLETITLE \cody{GAMESS}  \\[.75ex]
        size in mebibyte & 1071.8           & 951.7             & 642.7    & 512.0            & 381.5    &  601.52 & 180.7 & 306.19          \\[1ex]
        tuned GB/s       & 217.0             & 213.4            & 188.2             & 143.3           & 154.2     & 194.7 & 137.7 & 146.9         \\[1ex]
   best brute-force GB/s &  235.5           & 212.2              & 175.8            & 147.6            & 152.5   & 214.8 & 147.0 & 146.6          \\
 shared memory buffer size  & 5120             & 3584            & 5632           & 5632            & 5632   & 3072 & 5632 & 3584           \\
    \% diff. from tuned  & 8.5\%            & -0.5\%              & -6.5\%            & 3.0\%            & 1.1\%  & 10.3\% & 6.7\% & -0.2\% \\[1ex]
  worst brute-force GB/s &  157.2            & 159.5              & 117.8             & 92.1            & 92.6 & 131.5 & 88.6 & 88.9            \\
  shared memory buffer size  & 3072             & 7680             & 1024            & 1024           & 1024    & 2048 & 1024 & 1024          \\
    \% diff. from tuned  & 27.5\%             & 25.3\%              & 37.4\%             & 35.7\%           & 40.0\% & 32.4\% & 35.6\% & 39.5\%       \\[1ex]
  tuning speed GB/s & 2172.3 & 2126.8 & 1471.3 & 1087.0 & 745.4 & 1288.2 & 428.0 & 625.8 \\[1ex]
  tuned w/tuning overhead GB/s & 197.3 & 194.0 & 166.8 & 126.6 & 127.8 & 169.1 & 104.2 & 119.0 \\
  \% diff. from best case & 19.3\% & 9.4\% & 5.4\% & 16.6\% & 19.3\% & 27.0\% & 41.0\% & 23.2\%\\
  \% diff. from worst case & 20.3\% & 17.8\% & 29.4\% & 27.3\% & 27.5\% & 22.2\% & 14.9\% & 25.3\%\\
\end{tabular}%
}
    \caption{Comparison between our shared memory optimization and brute-force search for decoding and writing. The input quantization codes are generated by cuSZ with a relative error bound of $10^{-3}$. Negative percentages denote cases where our optimization outperformed the fastest brute-force case.}
    \label{tab:shmem-opt}
\end{table}

\begin{table*}[ht]
\tableSETUP
\resizebox{\textwidth}{!}{\begin{tabular}{@{} >{\scshape\bfseries}l *{3}{ *{8}{r}  } @{}}
                                &
    \TABLETITLE{HACC}           &
    \TABLETITLE{EXAALT}         &
    \TABLETITLE{CESM}           &
    \TABLETITLE{Nyx}            &
    \TABLETITLE{Hurr.}          &
    \TABLETITLE \cody{QMC.}  & 
    \TABLETITLE \cody{RTM}   & 
    \TABLETITLE \cody{GAMESS}   &
    \TABLETITLE{HACC}           &
    \TABLETITLE{EXAALT}         &
    \TABLETITLE{CESM}           &
    \TABLETITLE{Nyx}            &
    \TABLETITLE{Hurr.}          &
    \TABLETITLE \cody{QMC.}  & 
    \TABLETITLE \cody{RTM}   & 
    \TABLETITLE \cody{GAMESS}   &
    \TABLETITLE{HACC}           &
    \TABLETITLE{EXAALT}         &
    \TABLETITLE{CESM}           &
    \TABLETITLE{Nyx}            &
    \TABLETITLE{Hurr.}          &
    \TABLETITLE \cody{QMC.}  & 
    \TABLETITLE \cody{RTM}   & 
    \TABLETITLE \cody{GAMESS}
    \\[.5ex]
    {techniques}                &
    \multicolumn{8}{c}{\scshape
    original self-sync., gb/s}  &
    \multicolumn{8}{c}{\scshape
    optimized self-sync., gb/s} &
    \multicolumn{8}{c}{\scshape
        optimized gap array, gb/s}
    \\
    \cmidrule(lr){2-9}
    \cmidrule(lr){10-17}
    \cmidrule(l){18-25}
    {size in mebibyte}          &
    1071.8                      & 951.7  & 642.7  & 512.0  & 381.5  & 1071.8 & 601.5 & 180.7 & 306.2 & 951.7  & 642.7  & 512.0  & 381.5 & 601.5 & 180.7 & 306.2 & 1071.8 & 951.7  & 642.7  & 512.0  & 381.5  & 601.5 & 180.7 & 306.2        \\
    {compr. ratio}              &
    3.18                        & 2.40   & 9.60   & 15.99  & 9.78   & 2.46 & 8.41 & 12.10 & 3.18  & 2.40   & 9.60   & 15.99  & 9.92 & 2.45 & 8.62 & 12.45 & 3.00   & 2.26   & 9.04   & 15.05  & 9.33 & 2.31 & 8.12 & 11.71          \\
    \cmidrule(lr){2-9}
    \cmidrule(lr){10-17}
    \cmidrule(l){18-25}
    {intra-seq. sync.}          & 155.4  & 133.8  & 264.4  & 469.9  & 295.3 & 112.3 & 245.9 & 363.4 & 208.0  & 169.2  & 345.9  & 430.8  & 253.1  & 144.0 & 234.2 & 352.0 & -      & -      & -      & -      & -    & -      & -      & -     \\
    {inter-seq. sync.}          & 2295.3 & 1844.1 & 4240.8 & 6097.6 & 3436.6 & 1770.4 & 2086.2 & 3460.8 & 2145.6 & 1783.2 & 4813.2 & 4761.9 & 3443.0 & 1646.4 & 1838.4 & 3036.6 & -      & -      & -      & -      & -    & -      & -      & -     \\
    {get output idx.}           & 573.8  & 435.1  & 1516.0 & 2293.6 & 1487.7 & 439.3 & 1155.0 & 1765.1 & 569.0  & 432.3  & 1494.4 & 2189.1 & 1405.8 & 433.2 & 1060.6 & 1614.5 & 268.6  & 239.7  & 495.1  & 694.4  & 384.5 & 225.6 & 355.1 & 515.7 \\
    {tune shared mem.}          & -      & -      & -      & -      & -      & -      & -      & -     & 2172.3 & 2126.8 & 1471.3 & 1087.0 & 745.4  & 1180.0 & 406.5 & 683.0 & 2111.0 & 2128.8 & 1473.3 & 1234.6 & 741.8 & 1201.3 & 442.6 & 734.7\\
    {decode and write}          & 60.3   & 70.8   & 7.0    & 5.6    & 7.0  & 59.8 & 10.2 & 6.0 & 219.7  & 211.4  & 186.5  & 143.9  & 153.7 & 194.1 & 138.8 & 161.3 & 214.0  & 210.1  & 186.1  & 168.9  & 153.0 & 195.4 & 148.5 & 173.0 \\
    \cmidrule(lr){2-9}
    \cmidrule(lr){10-17}
    \cmidrule(l){18-25}
    {overall, decode}           & 39.7   & 40.9   & 6.8    & 5.5    & 6.8   & 35.1 & 9.6 & 5.9  & 83.0   & 71.5   & 101.9  & 92.1   & 78.1 & 63.1 & 64.8 & 87.3  & 112.8  & 106.4  & 123.9  & 122.4  & 95.4 & 96.3 & 84.7 & 110.1 \\
    {speedup}                   &
    \SPEEDUP{1.50×}             &
    \SPEEDUP{1.57×}             &
    \SPEEDUP{0.27×}             &
    \SPEEDUP{0.09×}             &
    \SPEEDUP{0.27×}             &
    \SPEEDUP{1.48×}             &
    \SPEEDUP{0.33×}             &
    \SPEEDUP{0.16×}             &
    \SPEEDUP{3.14×}             &
    \SPEEDUP{2.74×}             &
    \SPEEDUP{4.05×}             &
    \SPEEDUP{1.55×}             &
    \SPEEDUP{3.15×}             &
    \SPEEDUP{2.66×}             &
    \SPEEDUP{2.25×}             &
    \SPEEDUP{2.36×}             &
    \SPEEDUP{4.27×}             &
    \SPEEDUP{4.08×}             &
    \SPEEDUP{4.92×}             &
    \SPEEDUP{2.07×}             &
    \SPEEDUP{3.85×}             &
    \SPEEDUP{4.07×}             &
    \SPEEDUP{2.94×}             &
    \SPEEDUP{2.98×}
    \\
    \cmidrule(lr){2-9}
    \cmidrule(lr){10-17}
    \cmidrule(l){18-25}
\end{tabular}%
}
    \caption{A comprehensive evaluation of all proposed decoding solutions on V100: Huffman decoders using cuSZ quantization codes generated with a relative error bound of $10^{-3}$ on V100. GB/s is computed relative to the size of the generated quantization codes, i.e, half the dataset size.}
    \vspace{-2mm}
\label{tab:eval-default}
\end{table*}%

Table \ref{tab:shmem-opt} examines this shared memory optimization by comparing the decoding throughputs achieved by our optimization and by a brute-force search (in increments of 512 bytes from 1024 to 8192) for the optimal amount of shared memory to launch in a single grid (the test datasets will be described in \cody{\SEC} \ref{sec:eval}). According to the table, the throughputs of the decoding using this shared memory optimization on all of the datasets are within 10\% of the maximum throughput in the brute-force search.
Furthermore, some of the datasets tested performed better under the optimization than the maximum throughput achieved in the brute-force search; this is because different sections of the dataset have different compression ratios\footnote{Note that in order to achieve maximum throughput, different sections must be decoded with different amounts of shared memory.}. Additionally note the percent difference between the worst possible throughput and the optimized throughput; if shared memory is not used appropriately, the decoder may incur a performance penalty of up to 40\%. 

\cody{However, when one considers the overhead of the tuning mechanism itself, as is shown on the last rows of Table \ref{tab:shmem-opt}, this overhead ranges from 10.0\%$\sim$32.2\%. Note that this overhead is smaller on large datasets and vice versa. This is because tuning, in practice, takes approximately 220 microseconds on all datasets. Nevertheless, even with this tuning overhead, the decoder avoids a performance penalty of 23.1\% on average. However, note that on the 3D reverse time migration (RTM) dataset---the smallest dataset---the performance penalty avoided is only 14.9\%, while the performance loss compared to the best case achieved by brute force is 41.0\%. This is because the relatively constant runtime of tuning impacts smaller datasets more. Furthermore, several entries in Table \ref{tab:shmem-opt} suggest that one can just use an architecture-specific shared memory buffer size, like 5632 on the V100. However, on the EXAALT dataset, if a buffer size of 5632 is used, then decoding happens at 168.9 GB/s, which is 14.8\% slower than the tuned decoder, even when accounting for tuning overhead. Therefore, it is worth using our proposed tuning approach to find optimal buffer sizes for different datasets on different architectures.}

Note that in order to prevent a large reduction in occupancy caused by allocating too much shared memory, there is a threshold up to which 
we can allocate an amount of shared memory proportional to the compression ratio: this threshold is called $T_{high}$, as aforementioned. To attain $T_{high}$ for a particular GPU architecture, calculate the amount of shared memory required to attain at least 25\% occupancy, and divide that amount by $2048$ to obtain $T_{high}$. 
For example, on the Nvidia Tesla V100, shared memory usage must be under $16384$ bytes to attain that level of occupancy, so the corresponding value of $T_{high}$ is 8. 
If the compression ratio exceeds $T_{high}$, our experiments have demonstrated that 3584 symbols are an optimal size for the buffer in most situations on the V100 GPU.

\section{Performance Evaluation}
\label{sec:eval}
In this section, we present our experimental setup (including platforms, baselines, and datasets) and our evaluation results.

\subsection{Experiment Setup}\label{sub:eval-setup}

\subsubsection{Evaluation Platforms}\label{subeval-setup-:platform}
We conduct our experimental evaluation on the Bridges-2 supercomputer~\cite{brown2021bridges} at Pittsburgh Supercomputing Center (PSC), of which each GPU node is equipped with two Intel Xeon Gold 6248 CPUs and eight 32 GB NVIDIA Tesla V100 GPUs.

\subsubsection{Comparison Baselines}\label{sub:eval-setup-baseline}
We compare our optimized Huffman decoding with multiple baselines. Specifically, we compare our solution with (1) the original self-synchronization-based Huffman decoder~\cite{weissenberger2018massively}, (2) the original gap-array-based Huffman decoder~\cite{yamamoto2020gaparray}, and (3) cuSZ's na\"ive Huffman decoder~\cite{cusz2020}.  
Note that as the original gap-array-based Huffman decoder~\cite{gaparray} cannot be adapted to multi-byte inputs due to a bug, we estimate its performance by trimming each multi-byte quantization code to a single byte, considering most quantization codes are concentrated in the middle.

\begin{table}[ht]
\tableSETUP
    \resizebox{.9\linewidth}{!}{\begin{tabular}{@{} >{\bfseries\scshape}lrr @{}}
                                      & \TABLECAPTION datum size  & \TABLECAPTION \#fields    \\[-.3ex]
    \TABLETITLE datasets              &
    \TABLETITLE dimensions            &
    \TABLETITLE examples(s)                                                                   \\
    \TABLECAPTION cosmology           & \TABLECAPTION 1,071.75 MB & \TABLECAPTION 6 in total  \\[-.6ex]
    HACC                              & 280,953,867               &  xx, vx                     \\
    \TABLECAPTION molecular dynamics  & \TABLECAPTION 951.73 MB   & \TABLECAPTION 6 in total  \\[-.6ex]
    EXAALT                            & 2338$\times$106711        &  dataset2.x  \\
    \TABLECAPTION climate             & \TABLECAPTION 642.70 MB   & \TABLECAPTION 33 in total \\[-.6ex]
    CESM-ATM                          & 26$\times$1800$\times$3,600 & CLDICE, RELHUM              \\
    \TABLECAPTION cosmology           & \TABLECAPTION 512 MB      & \TABLECAPTION 6 in total  \\[-.6ex]
    Nyx                               & 512$\times$512$\times$512 & baryon\_density           \\
    \TABLECAPTION climate             & \TABLECAPTION 381.47 MB   & \TABLECAPTION 13 in total  \\[-.6ex]
    Hurricane               & 4$\times$100$\times$500$\times$500 & CLDICE, QRAIN \\
    \TABLECAPTION quantum circuits    & \TABLECAPTION 601.52 MB   & \TABLECAPTION 2 in total  \\[-.6ex]
    \cody{QMCPack}               & 115$\times$69$\times$69$\times$288 & einspline, einspline.pre \\
    \TABLECAPTION petroleum exploration    & \TABLECAPTION 180.73 MB   & \TABLECAPTION 1 in total (3600 snapshots)  \\[-.6ex]
    \cody{RTM}               & 449$\times$449$\times$235 & snapshot-1000 \\
    \TABLECAPTION quantum chemistry    & \TABLECAPTION 306.19 MB   & \TABLECAPTION 3 in total  \\[-.6ex]
    \cody{GAMESS}               & 80,265,168 & dddd, ffdd, ffff
\end{tabular}}
    \caption{Real-world \texttt{float}-type datasets used in the evaluation.}
    \label{tab:datasets}
\end{table}

\subsubsection{Test Datasets}\label{subs:eval-setup-dataset}
We conduct our evaluation and comparison based on
\cody{eight} typical 1D$\sim$4D real-world HPC simulation datasets, \cody{including six from} Scientific Data Reduction Benchmarks~\cite{sdrbench}: 1D HACC cosmology simulation~\cite{hacc}, 2D LAMMPS (part of the EXAALT ECP project) molecular dynamics simulation~\cite{lammps}, 3D CESM-ATM climate simulation~\cite{cesm-atm}, 3D Nyx cosmology simulation~\cite{nyx}, 4D Hurricane ISABEL simulation~\cite{hurricane}, \cody{and 4D QMCPack quantum simulation \cite{qmcpack}}. They have been widely used in much prior work~\cite{sz17,liang2018error,liang2019improving,zhao2020significantly,zhao2021optimizing,tian2021optimizing,tian2021revisiting,cusz2020,jin2020adaptive,use-case-Franck} and are good representatives of production-level simulation datasets. \cody{Additionally, we also evaluate two datasets that highlight our decoders' potential to be used as in-memory compressors as discussed in \SEC \ref{sec:intro}, including 3D RTM simulation data for petroleum exploration \cite{jin2021improving} and 1D GAMESS data for quantum chemistry simulation \cite{pastri}.} Each dataset includes multiple snapshots and diverse fields.  Table~\ref{tab:datasets} presents the test datasets in detail.
The data sizes per snapshot are 1.1 GB, 952 MB, 643 MB, 512 MB, \cody{381 MB, 602 MB, 181 MB, and 306 MB} for the above \cody{eight} datasets, respectively. \cody{Note that the datasets tested are over 100 MB in size. This is because larger snapshots are more likely to be found in scientific applications, especially in in-memory applications where the datasets to be compressed often take up a significant portion of the total available memory. However, as we have verified by truncating and decoding the HACC dataset, datasets as small as 10 MB can exhibit speedups over the baseline cuSZ decoder. In addition, the datasets tested are all single-precision data, because the current cuSZ only works with single-precision data. However, since our underlying optimizations work on Huffman decoding of multibyte symbols, our technique applies to double-precision data as well.}

\subsection{Experimental Results}\label{sub:eval-result}

\subsubsection{Huffman Decoding}\label{subs:eval-result-dec}

Table \ref{tab:cr-of-techniques} illustrates the compression ratios of our optimized decoders and the baselines on the test datasets. 
We note that the differences between the compression ratios of different methods are up to around 10\%. 
Thus, 
compression ratio is not the primary factor for choosing the most appropriate Huffman decoding approach; by comparison, throughput is more important. 
Note that, although the ``original gap-array'' row in Table \ref{tab:cr-of-techniques} refers to an 8-bit decoder, we double the compression ratio, so it can be used as a baseline for comparison with 16-bit decoders.

\begin{table}[htbp]
\tableSETUP
\resizebox{\linewidth}{!}{%
\begin{tabular}{@{} >{\bfseries\scshape}l rrrrrrrr@{}}
                         & \TABLETITLE HACC & \TABLETITLE EXAALT & \TABLETITLE CESM & \TABLETITLE Nyx  & \TABLETITLE Hurr. & \TABLETITLE \cody{QMC.} & \TABLETITLE \cody{RTM} & \TABLETITLE \cody{GAMESS} \\[.75ex]
        size in mebibyte & 1071.8           & 951.7              & 642.7           & 512.0            & 381.5            & 601.5           & 180.7            & 306.2             \\[.75ex]
        baseline cuSZ    & 3.20             & 2.40               & 9.06             & 15.64            & 9.78            & 2.46             & 8.41            & 12.10              \\
                         & \SPEEDUP{1.000×} & \SPEEDUP{1.000×}   & \SPEEDUP{1.000×} & \SPEEDUP{1.000×} & \SPEEDUP{1.000×} & \SPEEDUP{1.000×} & \SPEEDUP{1.000×} & \SPEEDUP{1.000×}  \\[.75ex]
        ori. self-sync   & 3.18             & 2.40               & 9.60             & 15.99            & 9.92             & 2.45             & 8.62            & 12.45             \\
                         & \SPEEDUP{0.996×} & \SPEEDUP{1.000×}   & \SPEEDUP{1.059×} & \SPEEDUP{1.022×} & \SPEEDUP{1.014×} & \SPEEDUP{0.998×} & \SPEEDUP{1.026×} & \SPEEDUP{1.029×} \\[.75ex]
        opt. self-sync   & 3.18             & 2.40               & 9.60             & 15.99            & 9.92             & 2.45             & 8.62            & 12.45              \\
                         & \SPEEDUP{0.996×} & \SPEEDUP{1.000×}   & \SPEEDUP{1.059×} & \SPEEDUP{1.022×} & \SPEEDUP{1.014×} & \SPEEDUP{0.998×} & \SPEEDUP{1.026×} & \SPEEDUP{1.029×} \\[.75ex]
        ori. gap-array*  & 3.11             & 2.60               & 9.42             & 15.51            & 9.68             & 2.41             & 8.43           & 12.10             \\
                         & \SPEEDUP{0.972×} & \SPEEDUP{1.079×}   & \SPEEDUP{1.040×} & \SPEEDUP{0.992×} & \SPEEDUP{0.990×} & \SPEEDUP{0.982×} & \SPEEDUP{1.002×} & \SPEEDUP{1.000×} \\[.75ex]
        opt. gap-array   & 3.00             & 2.26               & 9.04             & 15.05            & 9.33             & 2.31             & 8.12            & 11.71             \\
                         & \SPEEDUP{0.938×} & \SPEEDUP{0.941×}   & \SPEEDUP{0.997×} & \SPEEDUP{0.962×} & \SPEEDUP{0.954×} & \SPEEDUP{0.939×} & \SPEEDUP{0.965×} & \SPEEDUP{0.968×}
\end{tabular}
 }
  \caption{Compression ratio of \cody{eight} evaluated methods. The original gap-array-based method is of 8-bit symbols, so their compression ratios are doubled to provide a fair comparison.}
  \label{tab:cr-of-techniques}%
\end{table}%

On the other hand, Table \ref{tab:gbps-of-techniques} shows the throughput of each decoding method in GB/s. The average speedup of our optimized self-synchronization solution compared to the baseline (in this case cuSZ's decoder) is 
\cody{2.74}$\times$, and the average speedup of our optimized gap-array solution is 
\cody{3.64}$\times$.
Note that that the speedup over the original implementations of self-synchronization and gap-array solutions is more notable on high compression-ratio datasets.  
This is because the original implementations do not write out symbols to global memory in an efficient manner which is in turn exacerbated by the fact that high compression-ratio datasets have more symbols to be written out to memory.
This underscores the importance of the optimizations for efficient memory access and use of shared memory introduced in \SEC\ref{sub:decode-write}, especially when considering quantization codes generated by effective prediction methods.
Note further that the original gap-array solution, although its GB/s numbers are computed relative to 8-bit quantization codes, still achieves performance numbers that are greater than our optimized self-synchronization solution. Nevertheless, in addition to the practical reasons detailed above, that solution also exhibits the same performance issues on high compression-ratio datasets described earlier.

\begin{table}[htbp]
\tableSETUP
\resizebox{\linewidth}{!}{%
\begin{tabular}{@{} >{\bfseries\scshape}lrrrrrrrr@{}}
                          & \TABLETITLE HACC & \TABLETITLE EXAALT & \TABLETITLE CESM & \TABLETITLE Nyx & \TABLETITLE Hurr. & \TABLETITLE \cody{QMC.} & \TABLETITLE \cody{RTM} & \TABLETITLE \cody{GAMESS} \\[.75ex]
   size in mebibyte       & 1071.8           & 951.7             & 642.7          & 512.0           & 381.5           & 601.5           & 180.7            & 306.2               \\[.75ex]
   {baseline cuSZ}        & 26.4             & 26.1               & 25.2             & 59.2            & 24.8   & 23.7             & 28.8           &     37.0     \\
                          & \SPEEDUP{1.00×}  & \SPEEDUP{1.00×}    & \SPEEDUP{1.00×}  & \SPEEDUP{1.00×} & \SPEEDUP{1.00×}   & \SPEEDUP{1.00×}  & \SPEEDUP{1.00×} & \SPEEDUP{1.00×} \\[.75ex]
   {ori. self-sync}       & 39.7             & 40.9               & 6.8              & 5.5             & 6.8      & 35.1              & 9.6             & 5.9            \\
                          & \SPEEDUP{1.50×}  & \SPEEDUP{1.57×}    & \SPEEDUP{0.27×}  & \SPEEDUP{0.09×} & \SPEEDUP{0.27×} & \SPEEDUP{1.48×}  & \SPEEDUP{0.33×} & \SPEEDUP{0.16×}   \\[.75ex]
   {opt. self-sync}       & 83.0             & 71.5               & 101.9            & 92.1            & 78.1   & 63.1 & 64.8 & 87.3            \\
                          & \SPEEDUP{3.14×}  & \SPEEDUP{2.74×}    & \SPEEDUP{4.05×}  & \SPEEDUP{1.55×} & \SPEEDUP{3.15×} &  \SPEEDUP{2.66×}  & \SPEEDUP{2.25×} & \SPEEDUP{2.36×}   \\[.75ex]
   {ori. gap-array 8-bit} & 84.4             & 87.7               & 30.0             & 17.5            & 37.3  & 87.2 & 31.9 & 25.9             \\
                          & \SPEEDUP{3.20×}  & \SPEEDUP{3.36×}    & \SPEEDUP{1.19×}  & \SPEEDUP{0.30×} & \SPEEDUP{1.50×}  & \SPEEDUP{3.68×}  & \SPEEDUP{1.11×} & \SPEEDUP{0.70×}   \\[.75ex]
   {opt. gap-array}       & 112.8            & 106.4              & 123.9            & 122.4           & 95.4    & 96.3 & 84.7 & 110.1          \\
                          & \SPEEDUP{4.27×}  & \SPEEDUP{4.08×}    & \SPEEDUP{4.92×}  & \SPEEDUP{2.07×} & \SPEEDUP{3.85×}  & \SPEEDUP{4.07×}  & \SPEEDUP{2.94×} & \SPEEDUP{2.98×}   \\[.75ex]
\end{tabular}
}
    \caption{Decoding throughputs of \cody{eight} evaluated methods.}
  \label{tab:gbps-of-techniques}%
\end{table}%

Table \ref{tab:eval-default} illustrates more details of the original self-synchronization solution as well as our optimized self-synchronization and gap array solutions, breaking down the algorithms into multiple phases. 
The table shows the impact of our architectural optimizations on our optimized Huffman decoders in break down.
As for the architectural optimizations for the self-synchronization phase, we obtain average speedups of \cody{11\%} for the most expensive self-synchronization phase---intra-sequence self-synchronization. We can see that significant speedups of up to 34\% are shown on lower compression ratio datasets, where this phase is a more significant bottleneck in decoding. The decoding and writing phase of both our optimized solutions, which both use our customized decoder, perform on average \cody{7.1}$\times$ faster than the baseline decoder's Huffman decoding phase and \cody{15.6}$\times$ faster than the original self-synchronization based Huffman decoding and writing phase. Nevertheless, our customized Huffman decoder does decode at a somewhat reduced bandwidth at high compression ratios, but this is compensated by performance gains elsewhere while decoding a high compression-ratio dataset.


We note that our optimized decoder achieves the lowest speedup relative to the baseline (1.55$\times$ and 2.07$\times$) on the Nyx-quant dataset. 
This can be explained by examining both the dataset and the baseline cuSZ decoder:
(1) 
The Nyx dataset is extremely high-compressible, and the encoded Nyx dataset consists mostly of codewords of length one. Since cuSZ's decoder works one bit at a time, it is able to decode more codewords. 
(2) Since fewer threads run on cuSZ's coarse-grained decoder, it does not encounter the same issues with high compression-ratio input that other finer-grained decoders have.
As a result, the baseline cuSZ decoder has a relatively high performance on the Nyx dataset compared to the other datasets. 

\subsubsection{cuSZ Decompression}\label{tab:eval-result-cusz}

\begin{figure}[]
     \centering\sffamily\footnotesize
     \includegraphics[width=\linewidth]{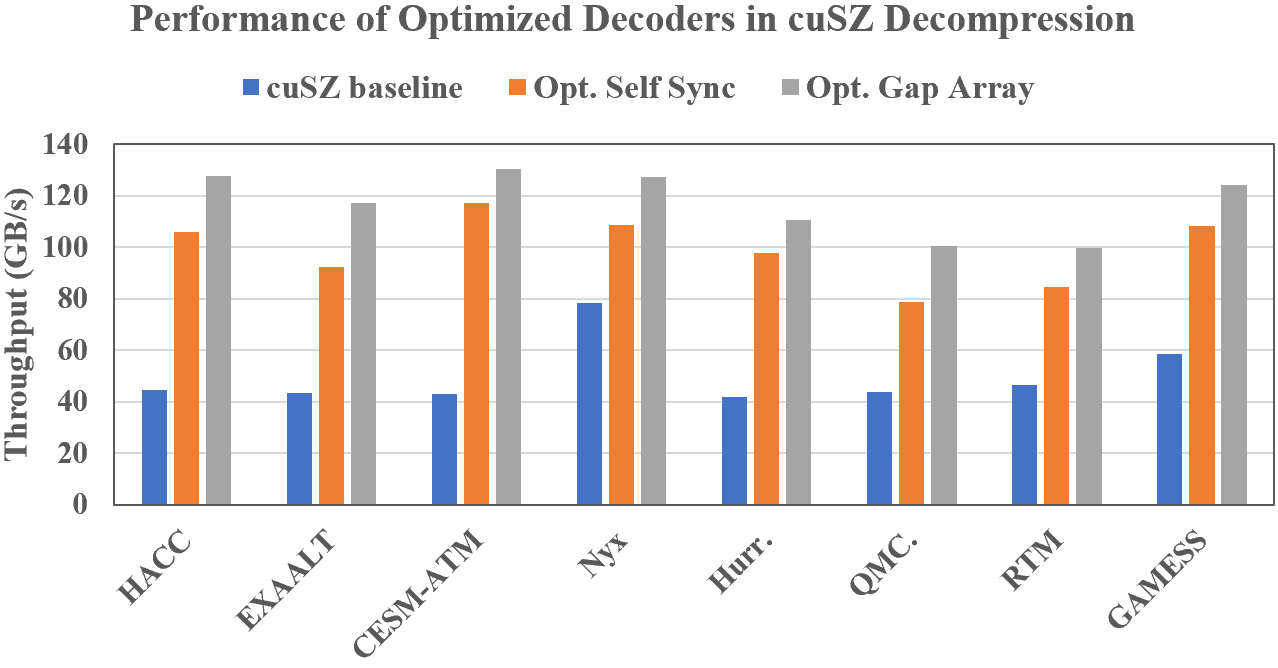}
   \caption{\cody{Performance comparison between our optimized decompression and the cuSZ baseline on V100 (with relative error bound $10^{-3}$). GB/s is computed relative to the size of the entire dataset.}}
     \label{fig:eval-cusz}
\end{figure}

Figure \ref{fig:eval-cusz} demonstrates the impact of our optimized decoders on the overall performance of cuSZ's decoder, by comparing the baseline decoder and our two optimized solutions. On average, substituting the baseline decoder with our optimized decoders resulted in \cody{2.08}$\times$ faster \cody{decompression} using self-synchronization and \cody{2.43}$\times$ faster \cody{decompression} 
using gap arrays. Note that in this scenario, we calculate GB/s with regard to the size of the scientific dataset itself rather than just the quantization codes. The reason that such a significant speedup can be attained is that cuSZ spends a substantial amount of time doing Huffman decoding; in the HACC dataset, cuSZ spent over 83\% of the overall decompression time doing Huffman decoding. As a result, with our optimized decoders, the optimized cuSZ can decode at speeds of over 100 GB/s on the V100 on most of the test cases.

\begin{figure}[]
     \centering\sffamily\footnotesize
     \includegraphics[width=\linewidth]{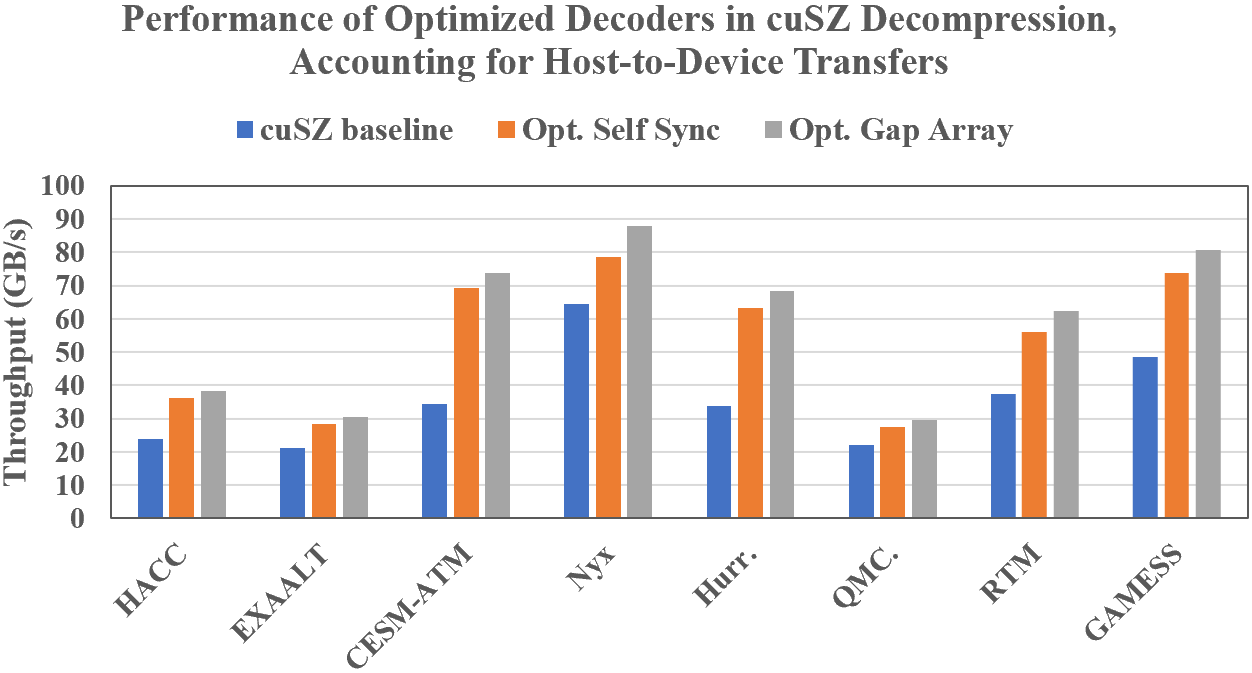}
   \caption{\cody{Performance comparison between our optimized decompression and the cuSZ baseline on V100 (with relative error bound $10^{-3}$), taking into account host-to-device memory transfers of compressed data. GB/s is computed relative to the size of the entire dataset.}}
     \label{fig:eval-cusz-plusmem}
\end{figure}

\cody{
Additionally, in many GPU applications, compressed data is retained on CPU memory, which is a larger resource than GPU memory. When data is needed for processing on the GPU, before decompression, compressed data must be transferred from CPU memory to GPU memory. Thus, in Figure \ref{fig:eval-cusz-plusmem}, we incorporate host-to-device ``memcpy'' into our evaluation. In this case, our optimized decompression performed, on average, 1.53$\times$ faster for self-synchronization and 1.65$\times$ faster for gap arrays over the cuSZ baseline. These speedups are lower than the results shown in Figure \ref{fig:eval-cusz} as data transfers are a bottleneck due to a relatively slow bandwidth between the GPU and the CPU. Further note that the datasets with a relatively high throughput are those with a high compression ratio; this is because there is less actual data being transferred, so the compressed data transfer is relatively fast for those datasets.
}

\subsection{Use-case of Our Two Decoders}\label{sub:use-case}

In this paper, we introduced two algorithms from the literature for fast parallel Huffman decoding and implemented deep optimizations for these two algorithms. Although both approaches are designed for fine-grained parallel Huffman decoding, and both approaches benefit from our architectural optimizations with regard to shared memory and decoding, both self-synchronization and gap array based parallel Huffman decoding are more suitable in some circumstances than others. 
Specifically, on one hand, if raw decoding performance is essential, our optimized gap array based Huffman decoding will inherently be faster than the self-synchronization based approach due to the costly and relatively unpredictable nature of finding synchronization points (particularly on GPUs). 
However, on the other hand, to obtain this raw decoding performance, applications must compute and store a gap array, which adds storage overhead as well as overhead to the encoder. Even in situations where these added costs are relatively insignificant, the encoder and the decoder must be coupled, meaning the encoder needs to be re-engineered. Therefore, in applications where flexibility is important, self-synchronization based Huffman decoding is more transparent to the encoders having different-source data and can balance this flexibility 

\section{Related Work}
\label{sec:related}

In addition to works focusing on parallel Huffman decoding that have been refered to extensively throughout the paper (namely, Weißenberger and Schmidt's work \cite{weissenberger2018massively}, Yamamoto \textit{et al.}'s work \cite{yamamoto2020gaparray}, and to a lesser extent Klein and Wiseman's work \cite{klein2003parhuff}), Johnston and McCreath additionally proposed an algorithm for massively parallel Huffman decoding \cite{johnston2017multicore}. Their algorithm proposes to deal with the problem of decoding variable length codes by starting decoding from every location in the bit sequence, eventually decoding the bit sequence correctly. Taking advantage of the GPU manycore architectures, the algorithm performs slightly faster on the GPU than a single-CPU-core Huffman decoder. This approach is not well-suited for our purposes, as their approach results in large amounts of computation for only marginal gains over CPU-based decoders. Thus, in this work we only consider the other two algorithms. 

Many works have been done that focus on optimizing parallel Huffman-type encoding. For example, Lal \textit{et al.} proposed a Huffman-based entropy encoding system (E\textsuperscript{2}MC) for GPUs \cite{Lal_Lucas_Juurlink_2017}. More recently, Tian \textit{et al.} proposed a fast parallel Huffman codebook construction algorithm and a parallel Huffman encoder for modern GPU architectures \cite{tian2021revisiting}. Since much work has already been focused on optimizing Huffman encoding, we do not presently consider optimizing encoding in our work.


\section{Conclusion}
\label{sec:conclusion}
In this work, we comprehensively analyze two state-of-the-art Huffman decoding algorithms for error-bounded lossy compression of scientific data and propose a deep architectural optimization for both algorithms. We also propose an efficient online approach to tune the shared memory to decode different parts of the data based on the data characteristics. We then adapt our optimized decoders to multi-byte data and integrate it into cuSZ. Our evaluation on \cody{eight} representative scientific datasets shows that our solution can improve cuSZ's Huffman decoding throughput by \cody{3.64}$\times$ on average and cuSZ's overall decoding throughput by \cody{2.43}$\times$ on average.
In the future, we plan to optimize and evaluate our Huffman decoder for generic datasets such as text data on Nvidia A100 GPU.


\section*{Acknowledgments}
\label{sec:acknowledgement}
\small This research was supported by the Exascale Computing Project (ECP), Project Number: 17-SC-20-SC, a collaborative effort of two DOE organizations---the Office of Science and the National Nuclear Security Administration, responsible for the planning and preparation of a capable exascale ecosystem, including software, applications, hardware, advanced system engineering and early testbed platforms, to support the nation's exascale computing imperative. The material was supported by the U.S. Department of Energy, Office of Science and Office of Advanced Scientific Computing Research (ASCR), under contract DE-AC02-06CH11357. This work was also supported by the National Science Foundation under Grants OAC-2003709, OAC-2034169, OAC-2042084, OAC-2104023, and OAC-2104023. 


\renewcommand*{\bibfont}{\small}
\printbibliography[]

\end{document}